\magnification\magstep1
\font\cst=cmr10 scaled \magstep3
\font\csc=cmr10 scaled \magstep2
\vglue 0.5cm

\centerline{\cst  Bouncing universes and their perturbations~:}
\vskip 0.5cm
 \centerline{\cst a simple model revisited}

\vskip 1 true cm
\centerline{\bf Nathalie Deruelle$^*$ {\rm and} Andreas Streich$^{**}$}
\vskip 0.5cm
\centerline{\it  $^*$Institut d'Astrophysique de Paris,}
\centerline{\it GReCO, FRE 2435 du CNRS,}
\centerline{\it 98 bis boulevard  Arago, 75014, Paris, France}

\centerline{and}

\centerline{\it Institut des Hautes Etudes Scientifiques,}
\centerline{\it 35 Route de Chartres, 91440, Bures-sur-Yvette, France}

\bigskip
\centerline{\it $^{**}$ Ecole polytechnique, 91128, Palaiseau, France}
\centerline{and}
\centerline{\it  ETH, Department of Computing Sciences,}
\centerline{\it  R\"amistrasse 101, CH-809, Zurich, Switzerland}

\medskip
\vskip 0.8cm
\centerline{30 April 2004}

\vskip 1cm
\noindent
{\bf Abstract}
\bigskip
We reconsider the toy model studied in [1] of a spatially closed Friedmann-Lema\^\i tre universe, driven by a massive scalar field, which deflates quasi-exponentially, bounces
and then enters a period of standard inflation. We find that the equations for the matter density perturbations can be solved
analytically, at least at lowest order in some ``slow-roll" parameter. We can therefore give, in that limit, the explicit spectrum of the post-bounce perturbations in terms of their
pre-bounce initial spectrum. Our result is twofold. {\it If} the pre-bounce growing and decaying modes are of the same order of magnitude at the bounce, then the spectrum of the
pre-bounce growing modes is carried over to the post-bounce {\sl decaying} modes (``mode inversion").  On the other hand, if, more likely, the pre-bounce growing modes dominate,
then resolution at next order indicates that their spectrum is nicely carried over, with reduced amplitude, to the post-bounce {\sl growing} modes.

\vfill\eject

\noindent
{\csc I. Introduction}
\medskip

 Four dimensional, general relativistic and  bouncing Friedmann-Lema\^\i   tre models, to which string-inspired ``pre-Big Bang" [2] and ``ekpyrotic" or ``cyclic" [3] universes could
reduce within some effective theory limit, have recently attracted renewed interest.  A still debated issue however is how the spectrum of initial, pre-bounce, matter
fluctuations is modified by the bounce (see e.g. [4]).

As a warm up exercise, a toy model of a bouncing universe was studied in [1] (see also [5])~:  a spatially closed Friedmann-Lema\^\i tre model, driven by a massive scalar
field, which deflates quasi-exponentially, bounces and then enters a period of standard inflation. Unfortunately, no definite prediction on the post-bounce spectrum of
perturbations was reached, the main reason being the singular behaviour of the evolution equation in the bounce region. 

In this Note, we reconsider this simple model and rewrite the evolution equation for the matter perturbations in a well-behaved form. Having done so, we are able to solve it
analytically, at least at lowest order in some ``slow-roll" parameter, that is, when the pre and post-bounce quasi-de Sitter periods are long enough.  We shall hence obtain the
explicit spectrum of the perturbations when they exit the Hubble radius during post-bounce inflation, in terms of their initial spectrum when they enter the Hubble radius during
pre-bounce deflation. 

As we shall see, two cases will arise. If the pre-bounce growing and decaying modes are of the same order of magnitude at the bounce, then ``mode mixing" will turn out to
reduce to mode inversion~: the pre-bounce spectrum of the cosmologically interesting growing modes is carried over to the post-bounce {\sl decaying} modes, and hence
 soon lost. (As for the post-bounce surviving growing modes, they inherit the pre-``big-bang" decaying mode spectrum, which, usually, is unfortunately blue, see [1-3].) 

 On the
other hand, if the pre-bounce decaying modes have become negligible at the bounce, then resolution at next order of the perturbation equation indicates that the  pre-bounce 
growing mode spectrum (usually scale invariant) is nicely carried over to the surviving, cosmologically interesting, post-bounce {\sl growing} modes with no modification, apart
from an overall reduction factor.

We shall conclude by a few remarks on the genericity of the result and the validity of the slow-roll approximation which was made to yield it.

\bigskip
\noindent
{\csc II. The background}
\medskip
Consider a spatially closed, homogeneous and isotropic universe with line element~: $ds^2=-dt^2+S^2(t)d\Omega^2_3$ where $t$ is cosmic time, $S(t)$ the scale factor and
$d\Omega^2_3$ the line element of a 3-dimensional unit sphere. If matter is just a scalar field $\varphi$ with mass $m$ the Einstein equations
reduce to the homogeneous Klein-Gordon and Friedmann-Lema\^\i tre equations~:
$$\ddot\varphi+3H\dot\varphi+m^2\varphi=0\qquad;\qquad 3\left(H^2+{1\over
S^2}\right)=\kappa\left({\scriptstyle{1\over2}}\dot\varphi^2+{\scriptstyle{1\over2}}m^2\varphi^2\right)\eqno(2.1)$$ where a dot
denotes differentiation with respect to cosmic time, where $H\equiv\dot S/S$ is the Hubble parameter and where $\kappa$ is Einstein's constant.

The system of equations (2.1) has been thoroughly studied, in particular in [6-8]. We shall retain here that there exist ranges of initial conditions for which
the scale factor has a minimum. We shall restrict ourselves to the case when such a bounce occurs, at $t=0$ without loss of generality, and set the initial conditions there~:
$\varphi(0)=\varphi_0$ and $\dot\varphi(0)=\dot\varphi_0$ (the initial condition for $S$ follows from the fact that $\dot S(0)=0$). Introducing  rescaled  initial condition, time, scalar field and scale factor as
$$\phi_0\equiv\sqrt{\kappa\over6}\,\varphi_0\quad,\quad \tau\equiv \phi_0m\,t\quad,\quad \phi\equiv{\varphi\over\varphi_0}\quad,\quad  a\equiv \phi_0m\,S
\,,\eqno(2.2)$$ 
as well as the auxiliary function $z(\tau)\equiv{1\over m}\sqrt{\kappa\over6}\,\dot\varphi$, the system (2.1)  becomes~:
$${d\phi\over d\tau}={z\over \phi_0^2}\quad,\quad {dz\over d\tau}=-{3z\over a}{da\over d\tau}-\phi\quad,\quad {da\over
d\tau}=\pm\sqrt{a^2\left(\phi^2+{z^2\over \phi_0^2}\right)-1}\eqno(2.3)$$ where the plus sign holds after the bounce ($\tau\geq0$) and the minus sign before. As for the
initial conditions, they  become
$$\phi(0)=1\quad,\quad z(0)=z_0\quad,\quad a(0)={1\over\sqrt{1+z_0^2/\phi_0^2}}\eqno(2.4)$$
(with $z_0\equiv{1\over m}\sqrt{\kappa\over6}\,\dot\varphi_0$).  As one can see from (2.3), the solution for $\tau<0$ corresponding to the set of initial
conditions $(\phi_0,z_0)$ can be obtained from the solution for $\tau>0$ corresponding to the set $(\phi_0,-z_0)$ by means of the transformation
$$a(\tau,z_0)=a(-\tau,-z_0)\ ,\  \phi(\tau,z_0)=\phi(-\tau,-z_0)\ ,\  z(\tau,z_0)=-z(-\tau,-z_0)\quad\phi_0\ \hbox{fixed}\,.\eqno(2.5)$$

Now, if the standard\footnote{$^1$}{See [7] for fine-tuned values of $\phi_0$ of order 1 which also yield inflation.
See [8] for a proof that all models recollapse. Generic values of $\phi_0$ of order 1 yield small universes which soon recollapse. Hence the toy model considered here is
inappropiate to describe a post-bounce universe which does not inflate and immediately enters a radiation era.} conditions for post-bounce inflation are imposed, that is if
$$\phi_0\gg 1\qquad\hbox{and}\qquad |z_0|\ll \phi_0\eqno(2.6)$$
then the solution of the system (2.3) can be approximated, at zeroth order in the ``slow-roll" parameter $1/\phi_0^2$, by (see
[1] who limited themselves to the case $z_0=0$)~:
$$a\simeq \cosh \tau\quad,\quad \phi\simeq 1\quad,\quad z\simeq {z_0\over\cosh^3\tau}
-{1\over3}\,{\sinh \tau\over\cosh^3 \tau}(\cosh^2\tau+2)\,.\eqno(2.7)$$
By comparison with direct numerical integration of (2.3), one sees that (2.7)  is a good approximation to the exact solution for $\phi$ and $a$ as long as
$|\tau|\ll
\phi_0$, and a good approximation for
$z$ on the much wider range
$|\tau|\ll
\phi_0^2$, that is as long as the universe is well within the two, pre and post-bounce, dust-like eras.\footnote{$^2$}{The analytical solution at next order in $1/\phi_0^2$ can
easily be obtained, see [1], but will not be used in this Note.}

\bigskip

\bigskip
\noindent
{\csc III. The evolution equation for the scalar perturbations}
\medskip 
We consider now the perturbed metric~: $ds^2=-(1+2\Phi)dt^2+S^2(t)(1-2\Psi)d\Omega^2_3$ and the perturbed scalar field $\varphi(t)+\chi$. In Fourier space, the
``scalar" perturbations $\Phi_n$, $\Psi_n$ and $\chi_n$ are functions of time and  of the Eigenvalues $n$ of the Laplacian on the $3$-sphere (defined as $\triangle
f_n=-n(n+2)f_n, n\in N$ and $n\geq2$). The $(kl)$, $(0k)$ and $(00)$ components of the linearized  Einstein equations then are, respectively (see, e.g. [9])~:
$$\eqalign{\Phi_n=\Psi_n\qquad,&\qquad \dot\Psi_n+H\Phi_n={\kappa\dot\varphi\over2}\chi_n\cr
-3H\dot\Psi_n-{k^2\Psi_n\over S^2}+{3K\Phi_n\over S^2}&={\kappa\over2}\left(\dot\varphi\dot\chi_n+\chi_n{dV\over
d\varphi}+2V\Phi_n\right)\cr}\eqno(3.1)$$
 where $k^2\equiv n(n+2)-3K$, with $K=1$, and where, in the toy model we consider here~: $V(\varphi)\equiv{\scriptstyle{1\over2}}m^2\varphi^2$. As is well-known [9], the
last equation can be rewritten, using the two constraints as~:
$$\ddot\Phi_n+\left(7H+{2\over\dot\varphi}{dV\over d\varphi}\right)\dot\Phi_n+\left[{k^2-5K\over S^2}+2\left(\kappa V+{H\over\dot\varphi}{dV\over
d\varphi}\right)\right]\Phi_n=0\,.\eqno(3.2)$$
Another form of that equation is easily found to be, see [9]~:
 $$ u^{\prime\prime}_n+\left[k^2-W(\eta)\right]u_n=0\quad\hbox{with}\quad u_n\equiv{a\over\varphi^\prime}\Phi_n\quad\hbox{and}\quad
W(\eta)=-{\varphi^{\prime\prime\prime}\over\varphi^\prime}+2{\varphi^{\prime\prime2}\over\varphi^{\prime2}}+\kappa{\varphi^{\prime2}\over2}\eqno(3.3)$$
where  a prime denotes differentiation with respect to conformal time $\eta$---related to cosmic time $t$ by $Sd\eta=dt$.

It is clear that none of the forms (3.1) (3.2) or (3.3) is suitable for integration when $\dot\varphi$ goes through zero (which is necessarily the case if there is to be quasi-de
Sitter regimes before and after the bounce, see e.g. equation (2.7) for $z\propto\dot\varphi$).\footnote{$^3$}{The authors of ref [1]  solved  the perturbation equation (3.2) (in the
particular case
$z_0=0$) using delicate  numerical matching techniques.} Now, it is easy to check that, at least when
$V(\varphi)\equiv{\scriptstyle{1\over2}}m^2\varphi^2$,\footnote{$^4$}{The result can easily be extended to any potential of the form~$V(\varphi)\equiv c_0+c_1\varphi^n$.}
 they can be put into the strictly equivalent, well-behaved form
$$\left\{\eqalign{{d (a\Phi_n)\over d\tau}&={z\over a^2}A_n\cr {dg_n\over d\tau}&={k^2\over a^5\phi}A_n+{\Phi_n\over
\phi_0^2}
\left(3-{k^2\over a^2\phi^2}\right)\cr}\right. \qquad\hbox{with}\quad A_n\equiv
a^3\left(zg_n-{k^2\Phi_n\over a^2\phi}\right)\,,\eqno(3.4)$$
where $a(\tau)$, $\phi(\tau)$ and $z(\tau)$ solve the background equations (2.3). (The first equation is nothing but the $(0k)$ linearized equation and the second is a rewriting of
the $(00)$ one in terms of the suitably chosen auxiliary function $g_n$.\footnote{$^5$}{In the
late-time dust-like era when
$\dot\varphi$ and $\varphi\propto \phi$ go periodically through zero, but $H$ remains positive, another well-behaved form is required and was given in [10].}) Once (3.4) is
solved and $\Phi_n$ and $g_n$ known, then the other scalar perturbations are given by 
$$\Psi_n=\Phi_n\qquad\hbox{and}\qquad\sqrt{3\kappa\over2}\,{\chi_n\over
\phi_0}={A_n\over a^3}=\left(zg_n-{k^2\Phi_n\over a^2\phi}\right)\,.\eqno(3.5)$$
  Note that, in view of the symmetry properties of the background solution, see (2.5), the solution of (3.4) is such that
$$\Phi_n(\tau,z_0,\Phi_n(0),g_n(0))=\Phi_n(-\tau,-z_0,\Phi_n(0),-g_n(0))\,.\eqno(3.6)$$

\bigskip
\noindent
{\csc IV. Relating the pre and post-bounce spectra at lowest order in $1/\phi_0^2$}
\medskip 
 
When the conditions (2.6) on the initial conditions are met, the term in $1/\phi_0^2$ in the system (3.4) can, at lowest order, be ignored. The
evolution equation for the scalar perturbations thus simplifies into
$${d(a\Phi_n)\over d\tau}={z\over a^2} A_n\quad,\quad {dg_n\over d\tau}\simeq{k^2\over a^5\phi}\,A_n\qquad\hbox{with, recall~:}\quad A_n\equiv
a^3\left(zg_n-{k^2\Phi_n\over a^2\phi}\right)\eqno(4.1)$$
where, at the same approximation,  the background functions $a(\tau)$, $\phi(\tau)$ and $z(\tau)$ are given by (2.7).
Differentiating $A_n$ once, one finds~: 
$${dA_n\over d\tau}\simeq-g_n\cosh^3\tau\,.\eqno(4.2)$$ Differentiating again, one gets the following, closed, equation for $A_n$ (or, equivalently for $\chi_n$)~:
$${d^2A_n\over d\tau^2}-3\tanh \tau{dA_n\over d\tau}+{k^2\over \cosh^2\tau}A_n\simeq0\,.\eqno(4.3)$$
 
Recalling that $k^2=(n-1)(n+3)$, the general solution of equation (4.3) is a sum of even and odd functions~:
$$A_n=\alpha_nA_n^{(1)}+\beta_nA_n^{(2)}\eqno(4.4)$$
where $(\alpha_n, \beta_n)$ are constants of integration, and

\medskip
\noindent
$\bullet$ for $\tau>0$~: 
$$A_n^{(1)}=\cosh^3\!\tau \, F\left[{\scriptstyle{-{n+2\over2}, {n\over2},-{1\over2}}}, {1\over\cosh^2\tau}\right]\quad,\quad A_n^{(2)}=F\left[{\scriptstyle{{1-n\over2},
{n+3\over2},{5\over2}}}, {1\over\cosh^2\tau}\right]\eqno(4.5)$$
$\bullet$ for $\tau<0$~:
$$A_n^{(1)}=\pm \cosh^3\!\tau \, F\left[{\scriptstyle{-{n+2\over2}, {n\over2},-{1\over2}}}, {1\over\cosh^2\tau}\right]\quad,\quad A_n^{(2)}=\mp
F\left[{\scriptstyle{{1-n\over2}, {n+3\over2},{5\over2}}}, {1\over\cosh^2\tau}\right]\eqno(4.6)$$
 where the upper {\it vs} lower signs hold for $n$ even {\it vs} $n$ odd, and where $F[{\scriptstyle{a,b,c}},x]$ is the
hypergeometric function (usually denoted $2F1[{\scriptstyle{a,b,c}},x]$).

The function $A_n$ being known, the scalar perturbation $\Phi_n$ follows from (4.1-2) and the approximate background solution (2.7). It reads~:
$$\eqalign{\Phi_n&=\alpha_n\Phi_n^{(1)}+\beta_n\Phi_n^{(2)}\qquad\qquad\hbox{with}\cr
\Phi_n^{(1,2)}&\simeq-{1\over(n-1)(n+3)}\,{1\over\cosh\tau}\,\left(A_n^{(1,2)}+z{dA_n^{(1,2)}\over
d\tau}\right)\cr}\eqno(4.7)$$
$$=-{1\over(n-1)(n+3)}\sqrt{x}\left\{A_n^{(1,2)}+{2\over3}{dA_n^{(1,2)}\over dx}\,x\,\sqrt{1-x}\,\left[\sqrt{1-x}(1+2x)\mp 3z_0x^{3/2}\right]\right\}
$$
where $x\equiv a^{-2}\simeq {1\over\cosh^2\tau}$ and where the upper {\it vs} lower sign hold for ($\tau>0$) {\it vs} ($\tau<0$). 
At the bounce~:
$$\eqalign{\Phi_{n|\tau=0}^{(1)}\simeq& -{1\over(n-1)(n+3)}\,\left[(n+1)\cos{n\pi\over2}+z_0\,n(n+2)\sin {n\pi\over2}\right]\cr
\Phi_{n|\tau=0}^{(2)}\simeq-&{3\over(n-1)(n+3)}\,\left[{\sin{n\pi\over2}\over
n(n+2)}-{z_0\over n+1}\cos{n\pi\over2}\right]\,,\cr}\eqno(4.8)$$
$$g_{n|\tau=0}^{(1)}=-n(n+2)\sin{n\pi\over2}\quad,\quad g_{n|\tau=0}^{(2)}={3\over n+1}\cos{n\pi\over2}\,.\eqno(4.9)$$
Writing down the
explicit expression of $\Phi_n$ in terms of hypergeometric functions is not particularly illuminating~: suffice it to say that it is a good approximation of the numerical solution
of the exact equations (2.3) (3.4) in the range $|\tau|\ll\phi_0$, and that it tends to constants for large $|\tau|$ (in practice $|\tau|$ bigger than a few unities) and
oscillates in the bouncing region more and  more as $n$ grows bigger.

\bigskip
We are now in a position to relate the pre and post-bounce spectra, at the order considered, that is the lowest in the slow-roll parameter $1/\phi_0^2$.

 Using the following asymptotic expansions of hypergeometric functions, 
$$F\left[{\scriptstyle{{1-n\over2},{n+3\over2},{5\over2}}}, x\right]=1+{\cal O}(x)\quad,\quad 
F\left[{\scriptstyle{-{n+2\over2}, {n\over2},-{1\over2}}}, x\right]=1+{n(n+2)\over2}x+{\cal O}(x^2)\,,\eqno(4.10)$$
the asymptotic behaviours of $\Phi_n$  (in practice for $|\tau|$ bigger than a few unities) are readily obtained from (4.7) (recalling that $x\simeq a^{-2}$)~:
$$\bullet\ \hbox{post-bounce region~:}\quad\Phi_n\sim G^n_{\rm post}+{D^n_{\rm post}\over a}\quad\hbox{with}\quad\left\{\eqalign{G^n_{\rm
post}&=-{\alpha_n\over3}\cr D^n_{\rm post}&=-{\beta_n+3z_0\,\alpha_n\over (n-1)(n+3)}\cr}\right.\eqno(4.11)$$
$$\bullet\ \ \hbox{pre-bounce region~:}\quad\Phi_n\sim D^n_{\rm pre\ }+{G^n_{\rm pre\ }\over a}\quad\hbox{with}\quad\left\{\eqalign{D^n_{\rm
pre}&=\mp {\alpha_n\over3}\cr G^n_{\rm pre}&=\pm{\beta_n+3z_0\,\alpha_n\over(n-1)(n+3)}\cr}\right.\eqno(4.12)$$
where the upper {\it vs} lower signs hold for even {\it vs} odd $n$.

At lowest order in the slow-roll parameter $1/\phi_0^2$, the pre and post-bounce spectra are thus very simply related~:
$$\left(\matrix{G^n_{\rm post}\cr D^n_{\rm post}}\right) =\left(\matrix{\ \ 0&\pm1\ \cr \mp1&0\cr}\right)\left(\matrix{G^n_{\rm pre}\cr D^n_{\rm pre}}\right)
+{\cal O}(1/\phi_0^2)\,.\eqno(4.13)$$
In the toy model and at the approximation considered here, ``mode-mixing" therefore reduces to a simple ``mode inversion", that is the pre-bounce
spectrum of the growing  modes is carried over to the post-bounce decaying modes (and vice-versa). In other words~: the pre-bounce spectrum of the cosmologically
interesting growing modes is carried over to the post-bounce {\sl decaying} modes, and hence
 soon lost. On the other hand, the post-bounce surviving growing modes,  inherit the pre-bounce decaying mode spectrum, which, usually, is unfortunately blue [1-3].

There is however a case when the result does not hold, to wit $\alpha_n=0$, that is when the pre-bounce decaying modes have become insignificant when reaching the
bouncing region. This is a case of physical interest, to which we now turn.

\bigskip
\noindent
{\csc V. Validity of the approximation and next order in $1/\phi_0^2$}
\medskip 
Let us first look at the structure of the ``tranfer matrix" at next order in the slow-roll parameter.  Including $1/\phi_0^2$ corrections will yield
$$\left(\matrix{G^n_{\rm post}\cr D^n_{\rm post}}\right) \simeq\left(\matrix{\ \ {c(n)/\phi_0^2}&\ \pm1\cr \mp1&{d(n)/\phi_0^2}\cr}\right)\left(\matrix{G^n_{\rm pre}\cr
D^n_{\rm pre}}\right)\,\eqno(5.1)$$
and will not change significantly the zeroth order result,  {\it if $D^n_{\rm pre}$ and $G^n_{\rm pre}$ are of the same order of magnitude}.
If, now, the initial conditions on $D^n_{\rm pre}$ and $G^n_{\rm pre}$ are such that 
$$D^n_{\rm pre}\ll 
{c(n)\over\phi_0^2}\, G^n_{\rm pre}\,,\eqno(5.2)$$ 
which, in view of (4.12),  is the case when $\alpha_n$ is vanishingly small so that the pre-bounce decaying modes have become insignificant when reaching the
bouncing region, then  the pre and post-bounce spectra become related by
$$G^n_{\rm post}\simeq {c(n)\over\phi_0^2}\, G^n_{\rm pre}\quad,\quad
D_{\rm post}^n\simeq \mp G_{\rm pre}^n\,.\eqno(5.3)$$
This is the physically relevant case studied numerically in [1] where $G^n_{\rm post}$ is found to be non zero even when $D^n_{\rm pre}=0$. The authors of [1] did not however
give the
$n$-dependance of the constant $c(n)$ and hence left open the question of how $G^n_{\rm post}$ was related to $ G_{\rm pre}^n$.
\medskip

The constant $c(n)$ can however be estimated as follows.

The exact equations for the perturbations are (3.4) where the background functions solve (2.3). In the previous Section we solved them at zeroth order in $1/\phi_0^2$, that is we
ignored the $1/\phi_0^2$ term in  (3.4) and used for the background functions the zeroth order approximation (2.7). To consistently iterate them at next order one should~:

1. keep the $1/\phi_0^2$ term in  (3.4), replacing $\Phi_n$ by the zeroth order solution~;

2. use for the  background functions the first order approximation of (2.3).

To estimate $c(n)$, we shall however ignore step 2, for the following reason~: treating the background at first order in the slow-roll parameter introduces logarithmic
corrections in the solutions of the perturbation equation which render more difficult the numerical extraction of the post-bounce growing mode. Of course, the difficulty can be 
overcome,\footnote{$^7$}{Indeed, the task of integrating numerically the background and the well-behaved perturbation equations (3.4) is in principle straightforward.} but is
perhaps not worth the effort as it seems a reasonable guess that the inclusion of the $1/\phi_0^2$ correction for the backgound should not affect the $n$-dependence of the
perturbation spectrum.

We therefore integrated numerically the set of equations (3.4) where, in the  $1/\phi_0^2$ term, we replaced $\Phi_n$  by the pre-bounce purely growing zeroth order solution,
that is $\Phi_n^{(2)}$, see (4.5-7), but where we used for the background functions the zeroth order approximation (2.7). 

We chose the initial conditions at the bounce, 
 given by (4.8-9) with $\alpha_n=0, \beta_n=1$.  With those initial conditions,  integration yields a solution $\Phi_n^{\rm iter}$ which, when $\phi_0^2\to\infty$,  is nothing
but the analytical solution $\Phi_n^{(2)}$ obtained in the previous Section, that is a mode which is exponentially growing before the bounce and exponentially decaying after the
bounce. For large but finite value for
$\phi_0^2$ on the other hand,  $\Phi_n^{\rm iter}$ no longer vanishes in the asymptotic regions but tends to small  constants, ($c^n_{\rm post},c^n_{\rm pre}$), which scale, as
they should, as
$1/\phi_0^2$. Let us therefore introduce the rescaled constants ($C^n_{\rm post}\equiv \phi_0^2\, c^n_{\rm post}$, $C^n_{\rm pre}\equiv \phi_0^2\, c^n_{\rm pre}$) which are
independent of
$\phi_0^2$.

Consider now the linear combination
$$\Phi_n^{\rm shooting}=\Phi_n^{\rm iter}\pm 3{C^n_{\rm pre}\over\phi_0^2}\Phi_n^{(1)}\eqno(5.4)$$
where the upper {\it vs} lower signs hold for $n$ even {\it vs} $n$ odd. Since $\Phi_n^{(1)}\to\mp 1/3$ in the pre-bounce  asymptotic region (see (4.12)), we have
$$\bullet\ \ \hbox{pre-bounce region~:}\quad\Phi_n^{\rm shooting}\to 0\eqno(5.5)$$
so that $\Phi_n^{\rm shooting}$ is the purely pre-bounce growing mode at the approximation considered. In the post-bounce region on the other hand (see (4.11))~:
$$\bullet\ \hbox{post-bounce region~:}\quad\Phi_n^{\rm shooting}\to {C^n_{\rm post}\mp C^n_{\rm pre}\over\phi_0^2}\eqno(5.6)$$

Therefore, from (4.12) and (5.3)~:
$$G^n_{\rm post}\simeq T_nG^n_{\rm pre}\quad\hbox{with}\quad T_n=\pm {k^2\over\phi_0^2}(C^n_{\rm post}\mp C^n_{\rm pre})\,.\eqno(5.7)$$

Numerical integration gives, pour each $n$ (and each value of the parameter $z_0$ entering the background zeroth order solution (2.7)) the values of the constants $C^n_{\rm
post}$ and
$C^n_{\rm pre}$ and it turns out, remarkably, that, for large $n$, $T_n$ does not depend on $n$. Hence the pre-bounce spectrum  encoded in $G^n_{\rm pre}$ is nicely carried over
to the post-bounce growing mode $G^n_{\rm post}$, although with an amplitude reduced by the overall factor $1/\phi_0^2$.

\bigskip
\noindent
{\csc VI. Concluding remarks}
\medskip 

In the very simple toy model of a bouncing universe that we studied in this paper,  and at lowest order in the slow-roll parameter, the spectrum of the pre-bounce
perturbations is carried over through the bounce with a simple (although unfortunate)  inversion of modes. That analysis however breaks down  if the pre-bounce decaying modes
are negligible in the pre-bounce region. In that case the analysis must be pushed to next order in the slow-roll parameter with the neat indication that the large $n$ post-bounce
{\sl growing} modes inherit without distortion the pre-bounce {\sl growing} mode spectrum.

These results do not depend on the value of $\dot\phi$ at the bounce. (Indeed numerical integration indicates that the value of $z_0$ only affects the overall amplitude of the
tranfer factor
$T_n$.)

 It would be surprising if they depended crucially on the particular potential ($V(\varphi)={1\over2}m^2\varphi^2$) chosen for the scalar field, as long as
there exist quasi-de Sitter regimes before and after the bounce. They should not be spoilt either when treating more carefully the logarithmic corrections, but that point
certainly deserves further attention.

Of course, it is not clear whether current string-inspired bouncing universes (``pre-Big-Bang", ``ekpyrotic" or ``cyclic") can fit such a simple framework. If not, toy models as
the one considered here become irrelevant.

\vskip 1cm
\noindent
{\bf Acknowledgements}. N.D. is grateful to the organisors of the stimulating cosmology workshop which took place in  Cambridge in July 2003, and thanks
in particular Neil Turok and Paul Steinhardt for very interesting discussions. 
\vskip 1cm
 
\noindent
{\csc References}
\medskip 

\item{[1]} C. Gordon, N. Turok, Phys. Rev. D {\bf 67} (2003) 123508

\item{[2]} G. Veneziano, Phys. Lett. B {\bf 265} (1991) 287;  M. Gasperini,  G. Veneziano, Phys. Rep. {\bf 373} (2003) (and references therein)

\item{[3]} J. Khoury, B.A. Ovrut, P.J. Steinhardt, N. Turok, Phys. Rev. D {\bf 64} (2001) 123522; J. Khoury, B.A. Ovrut, N. Seiberg, P.J. Steinhardt, N. Turok, Phys. Rev. D {\bf 65}
(2002) 086007; N. Turok, P.J. Steinhardt, Phys. Rev. D {\bf 65} (2002) 126003; J. Khoury, B.A. Ovrut, P.J. Steinhardt, N. Turok, Phys. Rev. D {\bf 66} (2002) 0406005 (and
references therein)

\item{[4]} R. Brustein, M. Gasperini, M. Giovaninni, V. Mukhanov, G. Veneziano, Phys. Rev. D {\bf 51} (1995) 6744 ;  M. Gasperini, N. Giovaninni, G. Veneziano, Phys. Lett. B {\bf 259}
(2003) 113; Brandenberger, F. Finelli, JHEP {\bf 0111} (2001) 056; R. F. Finelli, R. Brandenberger, Phys. Rev. D {\bf 65} (2002) 103522;  A.J. Tolley, N. Turok, hep-th/0204091; D.H.
Lyth, Phys. Lett. B {\bf 524} (2002) 1; J.C. Hwang, Phys. Rev. D {\bf 65} (2002) 063514; S. Tsujikawa, Phys. Lett. B {\bf 526} (2002) 179;  P. Peter, N. Pinto-Neto, Phys. Rev D {\bf
65} (2001) 123513 and Phys. Rev D {\bf 66} (2002) 063509; J. Martin, P. Peter, N. Pinto-Neto, D.J. Schwarz hep-th/0112128, 0204222, 0204227; J. C Fabris, R.G. Furtado,
P. Peter, N. Pinto-Neto, Phys. Rev D {\bf 67} (2003) 124003, P. Peter, N. Pinto-Neto, hep-th/0306005; J. Martin, P. Peter, Phys. Rev D {\bf 68} (2003) 103517; R. Durrer, F. Vernizzi,
hep-th/0203375~; L. Allen and D. Wands, astro-ph/0404441

\item{[5]} J. Hwang, H. Noh, Phys. Rev. D {\bf 65} (2002) 124010

\item{[6]}  H. Nariai, Prog. Theor. Phys. {\bf 46} (1971) 433;  H. Nariai, K. Tomita, 
Prog. Theor. Phys. {\bf 46} (1971) 776;  L. Parker, S.A. Fulling, Phys. Rev. D {\bf 7} (1973) 2357;  H. Nariai, Prog. Theor. Phys. {\bf
51} (1974) 613;  A.A. Starobinski,
Pisma A.J. {\bf 4} (1978) 155;  V.N. Melnikov, S,V. Orlov, Phys. Lett. A {\bf 70} (1979) 263; J.D. Barrow, R.A,  Matzner, Phys. Rev. D {\bf 21} (1980) 336; J. B. Hartle, S.W. Hawking,
Phys. Rev. D {\bf 28} (1983) 2960; S.W. Hawking, J.C. Luttrell, Nucl. Phys. B {\bf 247} (1984) 250; D.N. Page, Class. and Quant. Grav. {\bf 1} (1984) 417; V.A. Belinsky, L.P.
Grishchuk, Ya.B. Zel'dovich, I.M. Khalatnikov, J. Exp. Theor. Phys. {\bf 89} (1985) 346; M. Visser, Phys. Lett. B {\bf 349} (1995) 443; A. Yu Kamenshchik,  I.M. Khalatnikov, A. V.
Toporensky, Int. Journ. Mod. Phys. D {\bf 6} (1997) 673;  Int. Journ. Mod. Phys. D {\bf 7} (1998) 129; I.M. Khalatnikov, A. Yu Kamenshchik, Phys. Rep. {\bf 288} (1997) 513; C.
Molina-Paris, M. Visser, Phys. Lett. B {\bf 455} (1999) 90; S. Gratton, A. Lewis, N. Turok, Phys. Rev D {\bf 65} (2002) 043513

\item{[7]} N.J. Cornish, E.P.S. Shellard, Phys. Rev. Lett. {\bf 81} (1998) 3571

\item{[8]} H.J. Schmidt, Astron. Nachr. {\bf 311} (1990) 99

\item{[9]} J.M. Bardeen, Phys. Rev. D {\bf 22} (1980) 1882; M. Sasaki, Prog. Thor. Phys. {\bf 70} (1983) 394; V.F. Mukhanov, H.A. Feldman, R.H. Brandenberger, Phys. Rep. {\bf 215}
(1992) 203;  J.M. Stewart, Class. Quant. Grav. {\bf 7} (1990) 1169

\item{[10]} N. Deruelle, C. Gundlach, D. Polarski, Class. Quant. Grav, {\bf 9} (1992) 137

\end